\documentclass{article}
\usepackage{amsmath}
\usepackage{amsfonts}
\usepackage{amssymb}

\input{tcilatex}

\begin{document}

\title{Quantum deformations of associative algebras and integrable systems}
\author{ B.G.Konopelchenko \\
\\
Dipartimento di Fisica, Universita del Salento \\
and INFN, Sezione di Lecce, 73100 Lecce, Italy}
\maketitle

\begin{abstract}
\bigskip

Quantum deformations of the structure constants for a class of
associative noncommutative algebras are studied. It is shown that
these deformations are governed by the quantum central systems
which has a geometrical meaning of vanishing  Riemann curvature
tensor for Christoffel symbols identified with the structure
constants. A subclass of isoassociative quantum deformations is
described by the oriented associativity equation and, in
particular, by the WDVV equation. It is demonstrated that a wider
class of weakly (non)associative quantum deformations is connected
with the integrable soliton equations too. In particular, such
deformations for the three-dimensional and infinite-dimensional
algebras are described by the Boussinesq equation and KP
hierarchy, respectively.
\end{abstract}



\section{Introduction}
\setcounter{equation}{0}

Modern theory of deformations for associative algebras which was formulated
in the classical works by Gerstenhaber [1,2] got a fresh impetus with the
discovery of the Witten-Dijkgraaf-Verlinde-Verlinde (WDVV) equation [3,4].
Beautiful formalization of the theory of WDVV equation in terms of the
Frobenius manifolds given by Dubrovin [5,6] \ and its \ subsequent extension
to F-manifolds [7,8] have provided us with the remarkable realization (see
e.g. [5-11] ) of one of Gerstenhaber's approches to the deformation of
associative algebras which consists in the treatment of '' the set of
structure constants as parameter space for the deformation theory'' ( [1],
Chapter II, section 1 ). A characteristic feature of the theory of Frobenius
and F-manifolds is that the action of the algebra is defined on the tangent
sheaf of these manifolds [5-11].

A different method to describe deformations of the structure
constants for associative commutative algebra in a given basis has
been proposed recently in [12,13,14]. This approach consists 1) in
converting the table of
multiplication for an associative algebra in the basis \textbf{\ }$\mathbf{P}%
_{0},\mathbf{P}_{1},...,\mathbf{P}_{N-1},$ i.e.

\begin{equation}
\mathbf{P}_{j}\mathbf{P}_{k}=C_{jk}^{l}(x)\mathbf{P}_{l}, \qquad
j,k=0,1,...,N-1
\end{equation}
into the zero set $\Gamma $ of the functions
\begin{equation}
f_{jk}=-p_{j}p_{k}+C_{jk}^{l}(x)p_{l}, \qquad  j,k=0,1,...,N-1
\end{equation}%
with $p_{0},p_{1},...,p_{N-1}$ and deformation parameters $%
x^{0},x^{1},...,x^{N-1}$ being the Darboux canonical coordinates in the
symplectic space $R^{2N}$ and 2) in the requirement that the ideal J
generated by the functions $f_{jk}$ \ is the Poisson ideal, i.e.

\begin{equation}
\left\{ J,J\right\} \subset J
\end{equation}%
with respect to the standard Poisson bracket $\left\{ ,\right\} $ in $R^{2N}.
$ Here and below the summation over repeated index is assumed and this
index always run from 0 to N-1.

Deformations of the structure constants $C_{jk}^{l}$ defined by these
conditions are governed by the central system (CS) of equations consisting
from the associativity condition

\begin{equation}
C_{jk}^{l}(x)C_{lm}^{n}(x)-C_{mk}^{l}(x)C_{lj}^{n}(x)=0
\end{equation}
and the coisotropy condition

\begin{eqnarray}
\left[ C,C\right] _{jklr}^{m} & \doteqdot &
C_{sj}^{m}\frac{\partial C_{lr}^{s}}{%
\partial x^{k}}+C_{sk}^{m}\frac{\partial C_{lr}^{s}}{\partial x^{j}}%
-C_{sr}^{m}\frac{\partial C_{jk}^{s}}{\partial x^{l}}\allowbreak -C_{sl}^{m}%
\frac{\partial C_{jk}^{s}}{\partial x^{r}}+
\nonumber \\
& & + C_{lr}^{s}\frac{\partial C_{jk}^{m}}{\partial x^{s}}-
C_{jk}^{s}\frac{\partial C_{lr}^{m}}{\partial x^{s}}=0.
\end{eqnarray}

Such deformations of the structure constants have been called the
coisotropic deformations in [12-14]. CS (1.4), (1.5) is invariant under the
general transformations $x^{j}\rightarrow \widetilde{x}^{j}$ of the
deformation parameters \ with $C_{jk}^{l}$ being the (1,2) type tensor [14].
It has a number of other interesting properties. For \ the
finite-dimensional algebras this CS contains as the particular cases the
oriented associativity equation, the WDVV equation and certain
hydrodynamical \ type equations like the stationary dispersionless
Kadomtsev-Petviashvili (KP) equation ( or Khokhlov-Zabolotskaya equation)
[14]. For \ the infinite-dimensional polynomial algebras in the Faa' di
Bruno basis the coisotropic deformations are described by the universal
Whitham hierarchy of zero genus , in particular, by the dispersionless KP
hierarchy [12].

It was demonstrated in [14] that the theory of coisotropic deformations and
the theory of F-manifolds are essentially equivalent as far as the
characterization (1.4),(1.5) of the structure constants is concerned. \ One
of the advantages \ of the former is that it is formulated basically in a
simple framework of classical mechanics with the standard ingredients like
the phase space with the canonical coordinates $p_{j},x^{j}$ and the
constraints $f_{jk}=0$ which are nothing else than the Dirac's first class
constraints. This feature of the approach proposed in [12-14] strongly
suggests \ a way to built a natural and simple quantum version of
coisotropic deformations in parallel with the passage from the classical to
quantum mechanics.

In this paper \ we present the basic elements of the theory of quantum
deformations for a class of associative noncommutative algebras. A main idea
of the approach is \ to associate the elements of the Heisenberg algebra
with the elements \textsl{P}$_{j}$ of the basis for the algebra and
deformation parameters $x^{j}$ . Realising the table of multiplication ,
following the Dirac's prescription, as the set of equations selecting
''physical'' subspace in the infinite-dimensional linear space \ and
requireing that this subspace is not empty, one gets a system of equations,
the quantum central system (QCS),

\begin{equation}
\hbar \frac{\partial C_{jk}^{n}}{\partial x^{l}}-\hbar \frac{\partial
C_{kl}^{n}}{\partial x^{j}}+C_{jk}^{m}C_{lm}^{n}-C_{kl}^{m}C_{jm}^{n}=0
\end{equation}
which governs quantum deformations of the structure constants. Here $\hbar $
is the Plank's constant.

It is shown that a subclass of isoassociative quantum deformations, for
which the classical condition (1.4) is valid for all values of quantum
deformation parameters, is described by the oriented associativity equation
and , as the reduction, by the WDVV equation. A wider class of weakly
(non)associative quantum deformations is considered too. \ It is
characterised by nonvanishing quantum anomaly ( defect of associativity).
These deformations are also associated with integrable systems. It is shown
that for the three-dimensional algebra \ a class of such deformations is
described by the Boussinesq equation. For infinite-dimensional polynomial
algebras in the Faa' di Bruno basis \ the weakly (non)associative quantum
deformations \ of the structure constants are given by the KP hierarchy \ or
, more generally, by the multi-component KP hierarchy.

The paper is organised as follows. The definition of quantum deformations
and the derivation of the QCS (1.6) are given in the section 2.
Isoassociative quantum deformations and corresponding oriented associativity
equation are discussed in the section 3. In the next section 4 the weakly
(non)associativite deformations and an example of such deformation described
by the Boussinesq equation are considered. \ Quantum deformations of the
infinite-dimensional algebra and \ associated KP hierarchy are studied in
the section 5.2.

\bigskip

\section{Quantum deformations of associative algebras.}
\setcounter{equation}{0}

In the construction of quantum version of the coistropic deformations we
will follow basically the same lines as in the standard passage from the
classical mechanics to the quantum mechanics: subsitute a phase space by the
infinite-dimensional linear (Hilbert) space, introduce operators instead of
the canonically conjugated momenta and coordinates etc.

So, let \textit{A }be a N-dimensional associative algebra with ( or without)
unity element $\mathbf{P}_{0}$. We will consider a class of algebras which
posses a basis composed by pairwise commuting elements. Denoting elements of
a basis as $\mathbf{P}_{0},\mathbf{P}_{1},...,\mathbf{P}_{N-1}$ one writes
the table of multiplication
\begin{equation}
\mathbf{P}_{j}\mathbf{\ P}_{k}=C_{jk}^{l}(x)\mathbf{P}_{l},\qquad
 j,k=0,1,...,N-1
\end{equation}
where $x^{0},x^{1},...,x^{N-1}$ stand for the deformation parameters of the
structure constants. \ The commutativity of the elements of the basis
implies that $C_{jk}^{l}=C_{kj}^{l}.$

In order to define quantum deformations we first associate a set \ of linear
operators $\widehat{p}_{j}$ and $\widehat{x}^{j}$ ($j=0,1,...,N-1$) with the
elements $\mathbf{P}_{j}$ of the basis and the deformation parameters $x^{j}$
and require that these operators are elements of the Heisenberg algebra

\begin{equation}
\left[ \widehat{p}_{j},\widehat{p}_{k}\right] =0, \qquad \left[ \widehat{x}^{j},%
\widehat{x}^{k}\right] =0,\qquad \left[
\widehat{p}_{j},\widehat{x}^{k}\right] =\hbar \delta _{j}^{k},
\qquad  j,k=0,1...,N-1
\end{equation}
where $\hbar $ is the Planck's constant and $\delta _{j}^{k}$ is the
Kronecker symbol. \ The second step is to give a realization of the table of
multiplication (2.1) in terms of these operators. For this purpose we
introduce the set of operators $\widehat{f_{jk}}$ defined by

\begin{equation}
\widehat{f}_{jk}=-\widehat{p}_{j}\widehat{p}_{k}+C_{jk}^{l}(\widehat{x})%
\widehat{p}_{l}, \qquad  j,k=0,1,...,N-1.
\end{equation}

To simplify notations we will omit in what follows the label $\widehat{}$ in
the symbols of operators.

It is easy to see that the representation of the table of multiplication
(2.1) by the operator equations $f_{jk}=0$ is too restrictive. Indeed, it
implies the relation $\left[ f_{jk},p_{n}\right] =0$ which due to the
identity
\begin{equation}
\left[ p_{n},C_{jk}^{l}\right] =\hbar \frac{\partial C_{jk}^{l}}{\partial
x^{n}}
\end{equation}
gives $\frac{\partial C_{jk}^{l}}{\partial x^{n}}=0$. A right way
to quantize the first-class constraints from the classical
mechanics has been suggested long time ago by Dirac [15]. \ It
consists in the treatment of the first
-class constraints as the conditions selecting a subspace \textit{H}$%
_{\Gamma }$ of physical states in the Hilbert space \textit{H }by the
equations

\begin{equation}
f_{jk}\left| \Psi \right\rangle =0, \qquad  j,k=0,1,...,N-1
\end{equation}
where vectors $\left| \Psi \right\rangle \subset $\textit{H. }

The prescription (2.5) is the key point of the following

\textbf{Definition. }The structure constants $C_{jk}^{l}(x)$ are said to
define quantum deformations of an associative algebra if the operators \ $%
f_{jk}$ defined by (2.3) have a nontrivial common kernel.

If the conditions (2.5) are satisfied then any vector $\left| \Psi
\right\rangle $ belonging to \textit{H}$_{\Gamma }$ is invariant under the
group of transformations generated by operators $G=\exp (\alpha _{jk}f_{jk})$
where $\alpha _{jk}$ are parameters , i.e. $G\left| \Psi \right\rangle
=\left| \Psi \right\rangle .$ In such a form the above definition is the
quantum version of the classical condition (1.3).

The requirement (2.5) for the existence of the common eigenvectors with zero
eigenvalues for all opeators $f_{jk}$ imposes severe constraints on the
functions $C_{jk}^{l}(x).$ We begin with the well-known consequence of (2.5)
that is
\begin{equation}
\left[ f_{jk},f_{\ln }\right] \left| \Psi \right\rangle=0, \qquad
 j,k,l,n=0,1,...,N-1.
\end{equation}

This condition is the quantum version of the coisotropy condition $\left\{
f_{jk},f_{\ln }\right\} \mid _{\Gamma }=0$ in the classical case. \ Using
(2.2) and (2.4), one gets from (2.6) the relation

\begin{equation}
\left( \hbar ^{2}\frac{\partial ^{2}C_{jk}^{m}}{\partial x^{l}\partial x^{n}}%
-\hbar ^{2}\frac{\partial ^{2}C_{\ln }^{m}}{\partial x^{j}\partial x^{k}}%
-\hbar \left[ C,C\right] _{jk\ln }^{m}\right) p_{m}\left| \Psi \right\rangle
=0
\end{equation}
where the bracket $\left[ C,C\right] _{jk\ln }^{m}$ is defined in (1.5).

So, equations (2.6) are satisfied if%
\begin{equation}
\hbar \frac{\partial ^{2}C_{jk}^{m}}{\partial x^{l}\partial x^{n}}-\hbar
\frac{\partial ^{2}C_{\ln }^{m}}{\partial x^{j}\partial x^{k}}-\left[ C,C%
\right] _{jk\ln }^{m}=0, \qquad  j,k,l,n,m=0,1,...,N-1.
\end{equation}
This constraint is the quantum version of the cosisotropy condition (1.5).

To derive the quantum version of the associativity condition (1.4) we use
the identity
\begin{eqnarray}
 \left( p_{j}p_{k}\right) p_{l}-p_{j}\left( p_{k}p_{l}\right)
=&  p_{j}f_{kl}-p_{l}f_{jk}+C_{kl}^{m}f_{jm}-C_{jk}^{m}f_{lm}+ \quad &
\nonumber
\\
&  + \left(\hbar
\frac{\partial C_{jk}^{n}}{\partial x^{l}}-\hbar \frac{\partial C_{kl}^{n}}{%
\partial x^{j}}+C_{jk}^{m}C_{lm}^{n}-C_{kl}^{m}C_{jm}^{n}\right) p_{n}. &
\end{eqnarray}
It implies that
\begin{equation}
\left( \left( p_{j}p_{k}\right) p_{l}-p_{j}\left( p_{k}p_{l}\right) \right)
\left| \Psi \right\rangle =\left( \hbar \frac{\partial C_{jk}^{n}}{\partial
x^{l}}-\hbar \frac{\partial C_{kl}^{n}}{\partial x^{j}}%
+C_{jk}^{m}C_{lm}^{n}-C_{kl}^{m}C_{jm}^{n}\right) p_{n}\left| \Psi
\right\rangle
\end{equation}
for $\left| \Psi \right\rangle \subset \mathit{H}_{\Gamma }.$ Hence, if the
structure constants obey the equations

\begin{equation}
\hbar \frac{\partial C_{jk}^{n}}{\partial x^{l}}-\hbar \frac{\partial
C_{kl}^{n}}{\partial x^{j}}%
+C_{jk}^{m}C_{lm}^{n}-C_{kl}^{m}C_{jm}^{n}=0, \qquad
j,k,l,n=0,1,...,N-1
\end{equation}
then
\begin{equation}
\left( \left( p_{j}p_{k}\right) p_{l}-p_{j}\left( p_{k}p_{l}\right) \right)
\left| \Psi \right\rangle =0.
\end{equation}

We will refer to the conditions (2.12) and (2.11) as the quantum or weak
associativity conditions. We note that in virtue of (2.9) the request for
the ''strong'' associativity $\left( p_{j}p_{k}\right) p_{l}-p_{j}\left(
p_{k}p_{l}\right) =0$ implies $f_{jk}=0$ and \ so it is too rigid. Thus ,
similar to the table of multiplication (2.5) we require the fulfilment of
the associativity condition only on the ''physical'' subspace \textit{H}$%
_{\Gamma }$ instead of the whole linear space \textit{H . }

Equations (2.11) and (2.8) represent the quantum counterpart of the
classical CS (1.4),(1.5). These system of equations has in fact a much
simpler form since only part of them are independent. Indeed, one has the
following identity
\begin{equation}
\hbar T_{jk,\ln }^{m}=\hbar \frac{\partial R_{jlk}^{m}}{\partial x^{n}}%
-\hbar \frac{\partial R_{nlk}^{m}}{\partial x^{j}}%
-C_{js}^{m}R_{lkn}^{s}-C_{ns}^{m}R_{kjl}^{s}-C_{\ln
}^{s}R_{ksj}^{m}-C_{jk}^{s}R_{\ln s}^{m}-C_{lk}^{s}R_{sjn}^{m}
\nonumber
\end{equation}
where $T_{jk,\ln }^{m}$ denotes the l.h.s. of equation (2.8) and $%
R_{klj}^{n} $ stands for the l.h.s. of equation (2.11). Thus, we have

\textbf{Proposition 2.1 }\ The structure constants $C_{jk}^{l}(x)$ define a
quantum deformation of an associative algebra if they obey the equations
\begin{equation}
R_{klj}^{n}\doteqdot \hbar \frac{\partial C_{jk}^{n}}{\partial x^{l}}-\hbar
\frac{\partial C_{kl}^{n}}{\partial x^{j}}%
+C_{jk}^{m}C_{lm}^{n}-C_{kl}^{m}C_{jm}^{n}=0.\square
\end{equation}

We will refer to the system (2.13) as the quantum central system (QCS). We
emphasize that quantum deformations are defined in the category of
associative noncommutative algebras which possess commutative basis.

\textbf{\ Proposition 2.2 }\ If the structure constants define a quantum
deformation then

\begin{equation}
\left[ f_{jk},f_{lm}\right] =-\hbar K_{jk,lm}^{st}f_{st}, \qquad
j,k,l,m=0,1,...,N-1
\end{equation}
where
\begin{eqnarray}
K_{jk,lm}^{st}& = &
\frac{1}{2}\left( \delta _{m}^{t}\frac{\partial C_{jk}^{s}}{%
\partial x^{l}}+\delta _{l}^{t}\frac{\partial C_{jk}^{s}}{\partial x^{m}}%
-\delta _{k}^{t}\frac{\partial C_{lm}^{s}}{\partial x^{j}}-\delta _{j}^{t}%
\frac{\partial C_{lm}^{s}}{\partial x^{k}}+ \right. \nonumber \\
& & \quad \left. +
\delta _{m}^{s}\frac{\partial
C_{jk}^{t}}{\partial x^{l}}+\delta _{l}^{s}\frac{\partial C_{jk}^{t}}{%
\partial x^{m}}-\delta _{k}^{s}\frac{\partial C_{lm}^{t}}{\partial x^{j}}%
-\delta _{j}^{s}\frac{\partial C_{lm}^{t}}{\partial x^{k}}\right)
\end{eqnarray}

The proof is by direct calculation.

We note that the expession (2.15) exactly coincides with that which appear
in the coisotropic case for the Poisson brackets between the functions $%
f_{jk}$ . So, one has the same closed algebra for the basic objects $f_{jk}$
and $\widehat{f}_{jk}$ for the coisotropic and quantum deformations up to
the standard correspondence $\left[ ,\right] \longleftrightarrow -\hbar
\left\{ ,\right\} $ [15] between commutators and Poisson brackets.

The central systems (1.4), (1.5) and (2.13) which define coisotropic and
quantum deformations have rather different forms. In spite of this they have
some general properties in common. The invariance under the general
transformations of deformation parameters is one of them. \ Similar to the
coisotropic case [12-14] the quantum deformation parameters $x^{j}$ and
corresponding $p_{k}$ are strongly interrelated: they should obey the
conditions (2.2). So any change $x^{j}\rightarrow \widetilde{x}^{j}$ require
an adequate change $p_{k}\rightarrow \widetilde{p}_{k}$ in order the
relations (2.2) to be preserved. \ Thus, for the general transformation of
the deformation parametrs $x^{j}$ in our scheme one has
\begin{equation}
x^{j}\rightarrow \widetilde{x}^{j},p_{k}\rightarrow \widetilde{p}_{k}=\frac{%
\partial \widetilde{x}^{n}}{\partial x^{k}}p_{n}, \qquad
j,k=0,1,...,N-1.
\end{equation}
Note that the transformations (2.16) preserve the commutativity of the basis.

The requirement of the invariance for equations (2.5) readily implies that
\begin{equation}
C_{jk}^{l}(x)\rightarrow \widetilde{C}_{jk}^{l}(\widetilde{x})=\frac{%
\partial \widetilde{x}^{l}}{\partial x^{t}}\frac{\partial x^{s}}{\partial
\widetilde{x}^{j}}\frac{\partial x^{m}}{\partial \widetilde{x}^{k}}%
C_{sm}^{t}(x)+\hbar \frac{\partial \widetilde{x}^{l}}{\partial x^{m}}\frac{%
\partial ^{2}x^{m}}{\partial \widetilde{x}^{j}\partial \widetilde{x}^{k}}%
\end{equation}
under transformations (2.16). Then, it is a straightforward check that
equation (2.13) is also invariant. Hence, one has

\textbf{\ Proposition 2.3 }\ The QCS (2.13) is invariant under the general
transformations of the deformation parameters.

Furthermore, \ the relation (2.17) evidently coincides with the
transformation law \ of the Christoffel symbols and in the formula (2.13)
the tensor $R_{klj}^{n}$ is nothing but the Riemann curvature tensor
expressed in terms of the Christoffel symbols (see e.g. [16,17]).

Thus, we have

\textbf{Geometrical \ interpretation. }\ The QCS system (2.13) which governs
the quantum deformations in geometrical terms means the vanishing of the
Riemann curvature tensor $R_{klj}^{n}$ for the torsionless Christoffel
symbols $\Gamma _{jk}^{l}$ identified with the structure constants ($\Gamma
_{jk}^{l}=\hbar C_{jk}^{l}$).

In the standard terms  of the matrix-valued one-form  $\Gamma $ with
the matrix elements( see e.g. [17])
\begin{equation}
\Gamma _{k}^{l}=\left( C_{j}\right) _{k}^{l}dx^{j}=C_{jk}^{l}dx^{j}
\end{equation}
equation (2.13) looks like
\begin{equation}
\hbar d\Gamma +\Gamma \wedge \Gamma =0
\end{equation}
where $d$ and $\wedge $ denote the exterior differential and exterior
product, respectively. The flatness condition
\begin{equation}
\left[ \nabla _{j},\nabla _{l}\right] =0
\end{equation}
for the torsionless connection $\nabla _{j}=\hbar \frac{\partial }{\partial
x^{j}}+C_{j}$ is the another standard form of equation (2.13). In the
context of Frobenius manifolds the relation between the structure constants
and Christoffel symbols has been discussed within a different approach in
[6].

The identification of the structure constants with the Christoffel symbols
leads to certain constraints within such geometrical \ interpretation. For
instance, for an algebra with the unity element \textbf{P}$_{0}$, for which $%
C_{0k}^{l}=\delta _{k}^{l}$ , equation (2.13) immediately implies
\begin{equation}
\frac{\partial C_{jk}^{n}}{\partial x^{0}}=0, \qquad
j,k,n=0,1,...,N-1.
\end{equation}

Furthermore, if one requires that \textbf{P}$_{0}$ is invariant with respect
to the transformations (2.16) then $\frac{\partial \widetilde{x}^{j}}{%
\partial x^{0}}=\delta _{0}^{j}$ and $\frac{\partial x^{j}}{\partial
\widetilde{x}^{0}}=\delta _{0}^{j},j=0,1,...,N-1.$

For algebras with different properties ( semisimple, nilpotent etc) the
orbits generated by transformations (2.16) have quite different
parametrizations. \ For instance, for a semisimple algebra there is a basis
at which $C_{jk}^{l}=\delta _{jk}\delta _{j}^{l}$ (see e.g.[5,6]). \ Let us
denote the deformation parameters associated with this basis as $%
u^{0},u^{1},...,u^{N-1}.$ \ Then the corresponding orbit \ has the following
parametrization
\begin{equation}
C_{jk}^{l}(x)=\frac{\partial x^{l}}{\partial u^{m}}\frac{\partial u^{m}}{%
\partial x^{j}}\frac{\partial u^{m}}{\partial x^{k}}+\hbar \frac{\partial
x^{l}}{\partial u^{m}}\frac{\partial ^{2}u^{m}}{\partial x^{j}\partial x^{k}}%
\end{equation}
where $x^{m}(u),m=0,1,...,N-1$ are arbitrary functions. \ For a nilpotent
algebra for which all elemens have degree of nilpotency equal to two there
exists a basis at which all $C_{jk}^{l}=0.$ The general element of the
corresponding orbit is given by the formula
\begin{equation}
C_{jk}^{l}(x)=\hbar \frac{\partial x^{l}}{\partial u^{m}}\frac{\partial
^{2}u^{m}}{\partial x^{j}\partial x^{k}}
\end{equation}
where again $x^{m}(u)$ are arbitrary functions.

In the construction presented above we did not use concrete realization of
operators $p_{j}$ and $x^{k}$ . \ Any such realization provides us with a
concrete realization of the associative algebra under consideration and the
formulae derived . \ The most common representation of the Heisenberg
algebra (2.2) is given by the so-called Schrodinger representation at which
operators $\widehat{x}^{j}$ are the operators of multiplication by $x^{j}$ ,
$p_{j}$ are operators $\hbar \frac{\partial }{\partial x^{j}}$ and
wave-functions $\Psi (x)$ are elements of the space \textit{H} . \ In this
realization the associative algebra \textit{A }is the well-known algebra of
differential polynomials and equations (2.5) have the form

\begin{equation}
-\hbar \frac{\partial ^{2}\Psi }{\partial x^{j}\partial x^{k}}+C_{jk}^{l}(x)%
\frac{\partial \Psi }{\partial x^{l}}=0, \qquad  j,k=0,1,...,N-1.
\end{equation}

It is a simple check that the usual compatibility condition for the system
(2.24) (equality of the mixed third-order derivatives) is nothing else than
the conditions (2.12) and it is equivalent to equations (2.11). \ For an
algebra with the unity element one has $\hbar \frac{\partial \Psi }{\partial
x^{0}}=\Psi $ and , in virtue of (2.21) one has

\begin{equation}
\Psi (x)=e^{\frac{x^{0}}{\hbar }}\widetilde{\Psi }(x^{1},...,x^{N-1}).
\end{equation}

The system ( 2.24) is \ well-known in geometry. \ In the theory of the
Frobenius manifolds it is called the Gauss-Manin equation ( see e.g. [6,9]
). \ Such a system arises also \ in the theory of Gromov-Witten invariants
[18,19].

The standard quasiclassical \ approximation $\Psi =\exp (\frac{S(x)}{\hbar }%
),\hbar \rightarrow 0$ (see e.g. [15] ) performed for equations (2.24) give
rise to the Hamilton-Jacobi equations

\begin{equation}
-\frac{\partial S}{\partial x^{j}}\frac{\partial S}{\partial x^{k}}%
+C_{jk}^{l}\frac{\partial S}{\partial x^{l}}=0.
\end{equation}

These equations coincide with those for the generating function S for
Lagrangian submanifols which arise in the theory of coisotropic deformations
[14]. \ In this classical limit $\hbar \rightarrow 0$ the system (2.8),
(2.11) is reduced to the classical CS (1.4), (1.5) and the whole
construction presented above is reduced to that of coisotropic deformations.

Other realizations \ of the Heisenberg algebra (2.2) are of interest too.
Here we will mention only one of them given in terms of the standard
creation and annihilation operators $a^{+j}$ and $a_{j}$ and the Fock space.
The standard basis in he Fock space is given by the vectors

\begin{equation}
\left| n_{0},n_{1},...,n_{N-1}\right\rangle =\left(
n_{0}!n_{1}!...n_{N-1}!\right) ^{-\frac{1}{2}}\sqcap
_{k=0}^{N-1}\left( a^{+k}\right) ^{n_{k}}\left| 0\right\rangle
,n_{k}=0,1,2...
\nonumber
\end{equation}
where $a_{j}\left| 0\right\rangle =0,j=0,1,...,N-1$ and $\left\langle
0\right| 0\rangle =1.$

\noindent Then
\begin{equation}
\left| \Psi \right\rangle =\sum_{n_{k}=0}^{\infty
}A_{n_{0},n_{1},...,n_{N-1}}\left|
n_{0},n_{1},...,n_{N-1}\right\rangle \nonumber
\end{equation}
and the constraint (2.5) takes the form
\begin{equation}
\left( -a_{j}a_{k}+C_{jk}^{l}(a^{+})a_{l}\right) \left| \Psi
\right\rangle =0, \qquad j,k=0,1,...,N-1.
\end{equation}

This system of equations is equivalent to the infinite system of discrete
equations for the coefficients $A_{n_{0},...,n_{N-1}}$ while $%
C_{jk}^{l}(a^{+})$ obey QCS (2.13). \ Equations (2.27) \ define sort of
coherent states which could be relevant to the theory of quantum
deformations and its quasiclassical limit.

Finally, we note that several different '' quantization'' schemes for the
structures associated with the Frobenius manifolds , F-manifolds and
coisotropic submanifolds have been proposed in [18-22]. \ A comparative
analysis of these approaches and our scheme will be done elsewhere.

\section{Isoassociative quantum deformations and oriented
associativity equation}
\setcounter{equation}{0}

General quantum deformations described in the previous section contain as
a subclass \ of deformations for which the classical associativity condition
(1.4) is satisfied for all values of quantum deformation parameters. We will
refer to such deformations as isoassociative quantum deformations by analogy
with the isomonodromy and isospectral deformations. \ The formula (2.13)
implies

\textbf{\ Proposition 3.1 }\ Structure constants $C_{jk}^{l}(x)$ define
isoassociative quantum deformations of associative algebra if they obey the
equations

\begin{eqnarray}
C_{jk}^{m}C_{lm}^{n}-C_{kl}^{m}C_{jm}^{n} &=&0, \\
\frac{\partial C_{jk}^{n}}{\partial x^{l}}-\frac{\partial C_{kl}^{n}}{%
\partial x^{j}} &=&0.
\end{eqnarray}
In terms of the one-form $\Gamma $ (2.18) the system (3.1), (3.2) looks like
\begin{equation}
\Gamma \wedge \Gamma =0, \qquad  d\Gamma =0.
\end{equation}

Another way to arrive to the system (3.1), (3.2) consists in the treatment
of $\hbar $ in all the above formulae beginning with (2.2) not as the fixed
constant but as a variable parameter. In such interpretation the QCS (2.13)
from the very beginning splits into two equations (3.1) and (3.2), the
connection $\nabla _{j}$ from (2.20) becomes a pencil of flat torsionless
connection discussed in [5,6,8,9] and equations (2.24) coincide with the
Dubrovin's linear system for flat coordinates [5,6]. \ Thus, quantum
deformations for which $p_{j}$ and $x^{j}$ are elements of the pencil of
Heisenberg algebras (2.2) are of particular interest.

A way to deal with the system (3.1), (3.2) is to solve first equations
(3.2). They imply that
\begin{equation}
C_{jk}^{l}=\frac{\partial ^{2}\Phi ^{l}}{\partial x^{j}\partial x^{k}}
\end{equation}
where $\Phi ^{l},l=0,1,...,N-1$ are functions. Equation (3.1) then become
\begin{equation}
\frac{\partial ^{2}\Phi ^{m}}{\partial x^{j}\partial x^{k}}\frac{\partial
^{2}\Phi ^{n}}{\partial x^{m}\partial x^{l}}=\frac{\partial ^{2}\Phi ^{m}}{%
\partial x^{l}\partial x^{k}}\frac{\partial ^{2}\Phi ^{n}}{\partial
x^{m}\partial x^{j}}.
\end{equation}

The system (3.5) has appeared first in [5] (Proposition 2.3) as
the equation for the displacement vector. It has been rederived in
the different context in [23] and has been called the oriented
associativity equation there. \ In the form (3.3) it has \
appeared also in [24,25]. \ In our approach it describes the
isoassociative quantum deformation of the structure constants for
a class of associative noncommutative algebras. For this class of
deformations all operators $f_{jk}$ have a simple generating ''
function'', namely
\begin{equation}
\hbar ^{2}f_{jk}=\left[ p_{j,}\left[ p_{k},W\right] \right]
\nonumber
\end{equation}
where
\begin{equation}
W=-\frac{1}{2}\left( x^{m}p_{m}\right) ^{2}+\Phi ^{m}p_{m}.
\nonumber
\end{equation}

In the theory of coisotropic deformations [14] the deformations given by
equations (3.4) , (3.5) constitute a subclass of all deformations. So,
oriented associativity equation describes simultaneously both coisotropic
and isoassociative quantum deformations. In other words, one of the
characteristic features of the class of deformations given by the formulae
(3.1), (3.2) is that they remain unchanged \ in the process of
''quantization''. \ This means also that one can use both classical formulae
[14] and the quantum one ( previous section) to describe the properties of
these deformations. For instance, it was shown in [14] that in the natural
parametrization of the structure constants $C_{jk}^{l}$ by the eigenvalues
of the matrices $C_{j}$ and in terms of canonical coordinates $u^{j}$ the
system (3.2) becomes the system of conditions for the commutativity of N
hydrodynamical type systems. At the same time, the functions $\Phi ^{n}$
have a meaning of conserved densities for these \ hydrodynamical type
systems. All these results are valid for the isoassociative quantum
deformations too.

The oriented associativity equation (3.5) admits a well-known reduction to a
single superpotential F given by
\begin{equation}
\Phi ^{n}=\eta ^{nl}\frac{\partial F}{\partial x^{l}} \nonumber
\end{equation}
where $\eta ^{nl}$ is a constant metric. In this case equations
(3.5) becomes the famous WDVV\ equation [3,4]
\begin{equation}
\frac{\partial ^{3}F}{\partial x^{j}\partial x^{k}\partial x^{s}}\eta ^{st}%
\frac{\partial ^{3}F}{\partial x^{t}\partial x^{m}\partial x^{l}}=\frac{%
\partial ^{3}F}{\partial x^{l}\partial x^{k}\partial x^{s}}\eta ^{st}\frac{%
\partial ^{3}F}{\partial x^{t}\partial x^{m}\partial x^{j}}.
\end{equation}

Thus, the WDVV equation also describes the isoassociative quantum
deformations.

One more example of isoassociative quantum deformations is provided by the
Riemann space with the flat Hessian metric

$g_{jk}=\frac{\partial ^{2}\Theta }{\partial x^{j}\partial x^{k}}$
considered in [26] ( see also [27], Proposition 5.10). In this case [26]

\begin{equation}
C_{jk}^{l}=\hbar \Gamma _{jk}^{l}=\hbar g^{lm}\frac{\partial ^{3}\Theta }{%
\partial x^{j}\partial x^{k}\partial x^{m}},
\end{equation}
equation (3.2) is satisfied identically and the associativity
condition takes the form

\begin{equation}
\frac{\partial ^{3}\Theta }{\partial x^{j}\partial x^{k}\partial x^{s}}g^{st}%
\frac{\partial ^{3}\Theta }{\partial x^{t}\partial x^{m}\partial x^{l}}=%
\frac{\partial ^{3}\Theta }{\partial x^{l}\partial x^{k}\partial x^{s}}g^{st}%
\frac{\partial ^{3}\Theta }{\partial x^{t}\partial x^{m}\partial x^{j}}%
.
\end{equation}

This equation represents a rather nontrivial single-field reduction of
equation (3.5).

\section{Weakly (non)associative quantum deformations}
\setcounter{equation}{0}

All coisotropic deformations are isoassociative by construction [14]. \
For a subclass of them described by the oriented associativity equation the
CS is reduced to the system (3.1), (3.2). \ But, there is another subclass
of coisotorpic deformations for which the exactness conditions (3.2) are not
satisfied. \ For the finite-dimensional algebras such coisotropic
deformations are described by the stationary dispersionless KP equation and
other hydrodynamical type systems [14] . In the infinite-dimensional case
this type of deformations is described by the universal Whitham hierarchy of
zero genus and, in particular, by the dispersionless KP hierarchy [12].

What is the quantum version of coisotropic deformations of such type? \ One
naturally expects that they will not be the isoassociative one. On the other
hand, quantum deformations, for which equations (3.2) are not satisfied, are
governed by equation (2.13) with a nice geometrical meaning even if they are
not isoassociative.

All these suggest that the general quantum deformations defined by QCS
(2.13) without the additional exactness constraint (3.2) should be of
interest too. We will refer to such deformations as weakly associative or
weakly nonassociative quantum deformations.

The first term is due to the fact that according to (2.10) and (2.12) for
such deformations one has associativity for all values of quantum
deformation parameters \ not on the operator level, i.e. not on the whole
space\textit{\ H} , but only on the smaller ''physical'' subspace \textit{H}$%
_{\Gamma }$. The second name reflects the fact that the defect of
associativity
\begin{equation}
\alpha _{klj}^{n}\doteqdot C_{jk}^{m}C_{lm}^{n}-C_{kl}^{m}C_{jm}^{n}
\end{equation}
for such deformations is given by
\begin{equation}
\alpha _{klj}^{n}=-\hbar \left( \frac{\partial C_{jk}^{n}}{\partial x^{l}}-%
\frac{\partial C_{kl}^{n}}{\partial x^{j}}\right) .
\end{equation}

So, for small $\hbar $ or slowly varying structure constants the defect of
associativity is small.

For the matrix-valued two-form $\alpha _{q}$ with the matrix elements $%
\left( \alpha _{q}\right) _{k}^{n}\doteqdot \frac{1}{2}\alpha
_{klj}^{n}dx^{l}\wedge dx^{j}$ one has

\begin{equation}
\alpha _{q}=-\hbar d\Gamma
\end{equation}
where $\Gamma $ is defined in (2.18). One may refer to $\alpha _{q}$ also as
a quantum anomaly of associativity. Note that for an algebra with unity
element all elements $\alpha _{klj}^{n}$ with k or l or j =0 vanish.

Geometrical interpretation of the QCS (2.13) provides us with numerous
examples of weakly (non)associative quantum deformations. \ Any torsionless
flat connection gives us such deformation for certain associative algebra. \
In the generic case, for instance, these deformations are given by the
formulae (2.22) and (2.23) for the semisimple and nilpotent algebras,
respectively.

If there exists a metric $g_{jk}$ compatible with the Christoffel symbols $%
\Gamma _{jk}^{l}=\hbar C_{jk}^{l}$ \ then the generic deformation of the
structure constants is described by the formula

\begin{equation}
C_{jk}^{l}=\frac{1}{2}\hbar g^{nl}\left( \frac{\partial g_{nk}}{\partial
x^{j}}+\frac{\partial g_{jn}}{\partial x^{k}}-\frac{\partial g_{jk}}{%
\partial x^{n}}\right)
\end{equation}
where $g_{jk}$ is an arbitrary flat metric. Particular choice of the metric
gives us a specific deformation. For instance, for the diagonal flat metric $%
g_{jk}=\delta _{jk}H_{j}^{2}$ the weakly (non) associative \ deformations
are defined by the solutions of the well-known Lame system which describes
the orthogonal systems of \ coordinates in the N-dimensional Euclidean space
(see e.g. [16] ). \ For certain metrics , as, for example, for the Hessian
metric considered at the end of the previous section, the quantum anomaly
vanishes.

Other examples are provided by interpretation of the system (2.24) as the
system of equations for position vector in the affine differential geometry
( see e.g. [28]). We note also the papers [29,30] in which the equations
describing the geometry of submanifolds for a flat space have been reduced
to the WDVV type equations.

Different type of (non)associative deformations is given by the quantum
version of the coisotropic deformations of the finite-dimensional algebras
studied in [14]. As an illustrative example we will consider here the
three-dimensional (N=3) algebra with the unity element. \ The nontrivial
part of the table of multiplication is of the form

\begin{eqnarray}
\mathbf{P}_{1}^{2} &=&A\mathbf{P}_{0}+B\mathbf{P}_{1}+C\mathbf{P}_{2},
\nonumber \\
\mathbf{P}_{1}\mathbf{P}_{2} &=&D\mathbf{P}_{0}+E\mathbf{P}_{1}+G\mathbf{P}%
_{2}, \\
\mathbf{P}_{2}^{2} &=&L\mathbf{P}_{0}+M\mathbf{P}_{1}+N\mathbf{P}_{2}.
\nonumber
\end{eqnarray}

As in the paper [14] we consider the ''gauge'' B=0, C=1, G=0. \ The QCS
system (2.13) in this case assumes the form

\begin{eqnarray}
A+N-E &=&0, \nonumber \\
\hbar A_{x_{2}}-\hbar D_{x_{1}}+L-EA &=&0, \nonumber \\
-\hbar E_{x_{1}}+M-D &=&0, \\
\hbar D_{x_{2}}-\hbar L_{x_{1}}+ED-MA-ND &=&0,\nonumber \\
\hbar E_{x_{2}}-\hbar M_{x_{1}}+E^{2}-L-NE &=&0, \nonumber \\
-\hbar N_{x_{1}}+D-M &=&0 \nonumber
\end{eqnarray}
where $A_{x_{j}}\doteqdot \frac{\partial A}{\partial x^{j}}$ etc. This
system of equations implies that

\begin{eqnarray}
E &=&\frac{1}{2}A+\frac{3}{4}\epsilon ,N=-\frac{1}{2}A+\frac{3}{4}%
\varepsilon , \nonumber \\
L &=&\frac{1}{2}A^{2}+\frac{3}{4}\epsilon A-\hbar A_{x_{2}}+\hbar D_{x_{1}},
\\
M &=&D+\frac{\hbar }{2}A_{x_{1}} \nonumber
\end{eqnarray}
and
\begin{eqnarray}
A_{x_{2}}-\frac{4}{3}D_{x_{1}}+\epsilon _{x_{2}}-\frac{\hbar }{3}%
A_{x_{1}x_{1}} &=&0, \nonumber \\
D_{x_{2}}-\frac{3}{4}\left( A^{2}\right) _{x_{1}}-\frac{3}{4}\epsilon
A_{x_{1}}+\hbar A_{x_{1}x_{2}}-\hbar D_{x_{1}x_{1}} &=&0.
\end{eqnarray}
where $\epsilon (x_{2})$ is an arbitrary function. Eliminating D from the
system (4.8) ,one gets the equation

\begin{equation}
A_{x_{2}x_{2}}-\epsilon A_{x_{1}x_{1}}-\left( A^{2}\right) _{x_{1}x_{1}}+%
\frac{\hbar ^{2}}{3}A_{x_{1}x_{1}x_{1}x_{1}}+\epsilon _{x_{2}x_{2}}=0.
\end{equation}

At $\epsilon =$const it is the well-known Boussinesq equation which
describes surface waves (see e.g.[31] ).  This equation is integrable by
the inverse scattering transform method [32] similar to the famous
Korteweg-de Vries and KP equations (see e.g. [33,34,35]).

Equations (4.8) imply the existence of the function F such that

\begin{equation}
A=-\epsilon -2F_{x_{1}x_{1}},D=-\frac{3}{2}F_{x_{1}x_{2}}+\frac{\hbar }{2}%
F_{x_{1}x_{1}x_{1}}.
\end{equation}
In terms of the function F the system (4.8) or equation (4.9) become

\begin{equation}
F_{x_{2}x_{2}}-\epsilon F_{x_{1}x_{1}}+\frac{1}{2}\left( \epsilon
+2F_{x_{1}x_{1}}\right) ^{2}+\frac{\hbar ^{2}}{3}%
F_{x_{1}x_{1}x_{1}x_{1}}=0.
\end{equation}

The function $\tau $ defined by $F=\log \tau $ is the $\tau $- function for
the Boussinesq equation (4.9) and equation (4.11) is the Hirota equation to
it ( at $\epsilon =0$ see e.g. [36] ).

Any solution of the Boussinesq equation (4.9) or the Hirota equation (4.11)
provides us with the weakly (non)associative quantum deformation of the
algebra (4.5) with the structure constants given by the formulae (4.7),
(4.10). The quantum anomaly $\alpha _{q}$ (4.3) for these Boussinesq
deformations is of the form

\begin{equation}
\alpha _{q}=\hbar \left(
\begin{array}{ccc}
0 & \frac{1}{2}A_{x_{2}}+\frac{\hbar }{2}A_{x_{1}x_{1}} & \frac{1}{2}\left(
A^{2}\right) _{x_{1}} \\
0 & -A_{x_{1}} & -\frac{1}{2}A_{x_{2}}-\frac{\hbar }{2}A_{x_{1}x_{1}} \\
0 & 0 & A_{x_{1}}%
\end{array}%
\right) dx^{1}\wedge dx^{2}.
\end{equation}

To present a simple concrete example of deformation for the algebra (4.5) we
consider the following polynomial solution

\begin{equation}
F=\alpha \left( x^{1}\right) ^{2}+\beta x^{1}x^{2}+\left( \alpha \epsilon
-4\alpha ^{2}\right) \left( x^{2}\right) ^{2}+\gamma \left( x^{1}\right)
^{2}x^{2}+\frac{1}{3}\gamma \left( \epsilon -8\alpha \right) \left(
x^{2}\right) ^{3}-\frac{2}{3}\gamma ^{2}\left( x^{2}\right) ^{4}
\end{equation}
of the Hirota equation (4.11) with $\epsilon $=const where $\alpha ,\beta
,\gamma $ are arbitrary constants. \ This solution defines via (4.7) and
(4.10) the following weakly (non)associative deformation \ of the structure
constants

\begin{eqnarray}
A &=&-4\alpha -4\gamma x^{2},B=0,C=1, \nonumber \\
D &=&-\frac{3}{2}\beta -3\gamma x^{1},E=-2\alpha -2\gamma x^{2}+\frac{3}{4}%
\epsilon ,G=0, \nonumber \\
L &=&8\left( \alpha +\gamma x^{2}\right) ^{2}-3\epsilon \left( \alpha
+\gamma x^{2}\right) +\hbar \gamma ,M=-\frac{3}{2}\beta -3\gamma
x^{1}, \nonumber \\
N&=&2\alpha +2\gamma x^{2}+\frac{3}{4}\epsilon .
\end{eqnarray}
For this deformation the quantum anomaly is given by

\begin{equation}
\alpha _{q}=2\hbar \gamma \left(
\begin{array}{ccc}
0 & -1 & 0 \\
0 & 0 & 1 \\
0 & 0 & 0%
\end{array}%
\right) dx^{1}\wedge dx^{2}.
\end{equation}

General formulae (2.5) and (2.24) provide us with the auxiliary linear
problems for the Boussinesq equation. It is easy to show that equations
(2.5) for the algebra (4.5) with B=G=0, C=1 are equivalent to the following
two equations

\begin{eqnarray}
\left( p_{1}^{3}-\left( \frac{3}{2}A+\frac{3}{4}\epsilon \right)
p_{1}-\left( D+\hbar A_{x_{1}}\right) p_{0}\right) \left| \Psi \right\rangle
&=&0, \\
\left( p_{2}-p_{1}^{2}+Ap_{0}\right) \left| \Psi \right\rangle &=&0.
\end{eqnarray}

In the coordinate representation these equations look like

\begin{eqnarray}
\hbar ^{3}\Psi _{x_{1}x_{1}x_{1}}+\frac{3}{2}\hbar \left( u-\frac{\epsilon }{%
2}\right) \Psi _{x_{1}}+w\Psi &=&0, \\
\hbar \Psi _{x_{2}}-\hbar ^{2}\Psi _{x_{1}x_{1}}-u\Psi &=&0
\end{eqnarray}
where $u=-A$ and $w=-D-\hbar A_{x_{1}}$. Equations (4.18)-(4.19)
are the well-known auxiliary linear problems for the Boussinesq
equation at zero value of the spectral parameter [32-34].

Another set of linear problems one can obtains from the conditions (2.20).
For the Boussinesq algebra (4.5) the matrices $C_{1}$ and $C_{2}$ are

\begin{eqnarray}
C_{1} &=&\left(
\begin{array}{ccc}
0 & A & D \\
1 & 0 & \frac{1}{2}A+\frac{3}{4}\epsilon \\
0 & 1 & 0%
\end{array}%
\right) , \\
C_{2} &=&\left(
\begin{array}{ccc}
0 & D & \frac{1}{2}A^{2}-\hbar A_{x_{2}}+\hbar
D_{x_{1}}+\frac{3}{4}\epsilon A
\\
0 & \frac{1}{2}A+\frac{3}{4}\epsilon & D+\frac{\hbar }{2}A_{x_{1}} \\
1 & 0 & -\frac{1}{2}A+\frac{3}{4}\epsilon%
\end{array}%
\right) .
\end{eqnarray}

The commutativity condition (2.20) for the connection $\nabla _{j}=\hbar
\frac{\partial }{\partial x^{j}}+C_{j},J=1,2$ represents the compatibilty
condition for the linear problems

\begin{equation}
\left( \hbar \frac{\partial }{\partial x^{1}}+C_{1}\right) \varphi =0,\left(
\hbar \frac{\partial }{\partial x^{2}}+C_{2}\right) \varphi =0
\end{equation}
where $\varphi $ is the column $\varphi =\left( \varphi _{1,}\varphi
_{2},\varphi _{3}\right) ^{T}.$ Equations (2.20) for $C_{1}$ and $C_{2}$
given by (4.20) and (4.21) are equivalent to equations (4.7), (4.8), i.e. to
the Boussinesq equation. So, equations (4.22) represent the auxiliary matrix
linear problems for the Boussinesq equation. Equations (4.22) imply the
scalar equations

\begin{eqnarray}
\hbar ^{3}\varphi _{3,x_{1}x_{1}x_{1}}-\hbar \left( \frac{3}{2}A+\frac{3}{4}%
\epsilon \right) \varphi _{3,x_{1}}+\left( D+\frac{\hbar }{2}%
A_{x_{1}}\right) \varphi _{3} &=&0, \nonumber \\
\hbar \varphi _{3,x_{2}}+\hbar ^{2}\varphi _{3,x_{1}x_{1}}-A\varphi _{3}
&=&0.
\end{eqnarray}

Equations (4.23) are formally adjoint to equations (4.18)-(4.19) and their
compatibility condition gives rise to the same Boussinesq equation (4.9).

We would like to note that the ''zero curvature'' representation $\left[
\frac{\partial }{\partial x^{1}}+U,\frac{\partial }{\partial x^{2}}+V\right]
=0$ with the matrix-valued functions U and V is quite common in the theory
of the 1+1-dimensional integrable systems ( see e.g. [33-35]). The
particular representation of the form (4.22) is \ of interest at least by
two reasons. First, for instance, for the Boussinesq equation the elements
of the matrices $C_{1}$ and $C_{2}$ (4.20), (4.21) really coincide with the
components of the Christoffel symbol. \ Second, in such a representation the
elements of $C_{1}$ and $C_{2}$ are nothing else than the structure
constants of the deformed associative algebra (4.5).

All the above formulae for the Boussinesq \ quantum deformations in the
formal limit $\hbar \rightarrow 0$ ( with $\Psi =\exp \left( \frac{S}{\hbar }%
\right) $ ) are reduced to those for the coisotropic deformations of the
same algebra (4.5) which are described by the stationary dispersionless KP
equation [14].

Finally, we note that eliminating $F_{x_{1}x_{1}\text { }}$from the Hirota
equation (4.11) with the use of its  differential consequences, one
obtaines the equation

\begin{equation}
F_{x_{2}x_{2}x_{2}}F_{x_{1}x_{1}x_{1}}-F_{x_{2}x_{2}x_{1}}F_{x_{1}x_{1}x_{2}}=%
\frac{\hbar ^{2}}{3}\left(
F_{x_{1}x_{1}x_{2}}F_{x_{1}x_{1}x_{1}x_{1}x_{1}}-F_{x_{1}x_{1}x_{1}}F_{x_{1}x_{1}x_{1}x_{1}x_{2}}\right)
\nonumber
\end{equation}
which represents the ''quantum'' version of the Witten's equation [3]

\begin{equation}
F_{x_{2}x_{2}x_{2}}F_{x_{1}x_{1}x_{1}}-F_{x_{2}x_{2}x_{1}}F_{x_{1}x_{1}x_{2}}=0,
\nonumber
\end{equation}
i.e. equation (3.6) for the two-dimensional algebra without unity element
and the metric $\eta =\left(
\begin{array}{cc}
0 & 1 \\
1 & 0%
\end{array}%
\right) .$

\section{Quantum deformations of the infinite-dimensional algebra and
KP hierarchy.}\setcounter{equation}{0}

We will consider an infinite-dimensional algebra of polynomials generated by
a single element. In the so-called Faa' di Bruno basis the structure
constants of this algebra have the form [12]

\begin{equation}
C_{jk}^{l}=\delta _{j+k}^{l}+H_{j-l}^{k}+H_{k-l}^{j}, \qquad
j,k,l=0,1,2...
\end{equation}
where $H_{l}^{k}=0$ at $l\leq 0$ and $H_{l}^{0}=0.$ \ Coisotropic
deformations \bigskip of the structure constants (5.1) have been studied in
[12]. It was shown that they are described by the dispersionless KP
hierarchy.

Here we will discuss the quantum deformations of the same set (5.1) of the
structure constants.

\textbf{Proposition 5.1.} For the structure constants (5.1) the QCS (2.13)
is equivalent to the system
\begin{eqnarray}
\hbar \frac{\partial H_{j}^{k}}{\partial x^{l}}
+H_{j+k}^{l}+H_{j+l}^{k} - H_{j}^{k+l} +
\qquad \qquad \qquad\qquad \qquad \qquad \qquad  \nonumber \\
\qquad \qquad
+ \sum_{n=1}^{j-1}H_{j-n}^{k}H_{n}^{l} -
\sum_{n=1}^{l-1}H_{l-n}^{k}H_{j}^{n}-%
\sum_{n=1}^{k-1}H_{k-n}^{l}H_{j}^{n}=0, \qquad \nonumber \\
 j,k,l=0,1,2,.... &
\end{eqnarray}

Proof. Substitution of (5.1) into (2.13) gives

\begin{eqnarray}
\hbar \frac{\partial H_{j-m}^{k}}{\partial x^{l}}+\hbar \frac{\partial
H_{k-m}^{j}}{\partial x^{l}}-\hbar \frac{\partial H_{l-m}^{k}}{\partial x^{j}%
}-\hbar \frac{\partial H_{k-m}^{l}}{\partial x^{j}} &+& \nonumber \\
+ \sum_{n=0}^{\infty
}\left( \delta _{j+k}^{n}+H_{j-n}^{k}+H_{k-n}^{j}\right) \left( \delta
_{n+l}^{m}+H_{n-m}^{l}+H_{l-m}^{n}\right) &-& \nonumber \\
-\sum_{n=0}^{\infty }\left( \delta
_{l+k}^{n}+H_{l-n}^{k}+H_{k-n}^{l}\right) \left( \delta
_{n+j}^{m}+H_{n-m}^{j}+H_{j-m}^{n}\right)& = & 0 .
\end{eqnarray}
At j\TEXTsymbol{>}m,k\TEXTsymbol{<}m,l\TEXTsymbol{<}m using of the identity

\begin{equation*}
\sum_{p=n-1}^{k-1}H_{k-p}^{j}H_{p-n}^{m}=%
\sum_{p=n-1}^{k-1}H_{k-p}^{m}H_{p-n}^{j},
\end{equation*}
one gets equation (5.2) with the substitution j$\rightarrow $j-m. At \ m%
\TEXTsymbol{>}j,m\TEXTsymbol{>}l and m\TEXTsymbol{<}k equation (5.3) is
reduced to

\begin{equation}
\frac{\partial H_{k}^{j}}{\partial x^{l}}-\frac{\partial H_{k}^{l}}{\partial
x^{j}}=0.
\end{equation}

It is easy to see that equation (5.2) directly implies (5.4) due to the
symmetry of the nondifferential part in the indices k and l. \ An analysis
of all other choices of indices in (5.3) shows that the resulting equations
are all equivalent to (5.2) $\square .$

Solutions of the QCS (5.2) provide us with the quantum \ deformation of the
polynomial algebra in the Faa' di Bruno basis. \ In general, these
deformations are weakly (non)associative \ one and the quantum anomaly is
given by

\[
\alpha _{klj}^{n}=\hbar \left( \frac{\partial H_{l-n}^{k}}{\partial x^{j}}-%
\frac{\partial H_{j-n}^{k}}{\partial x^{l}}\right)
\]
or
\begin{equation}
\alpha _{q}=-\hbar dA
\end{equation}
where the matrix-valued one -form A has elements $\left( A\right)
_{k}^{n}=\sum_{l=n+1}^{\infty }H_{l-n}^{k}dx^{l}.$ At $\hbar \rightarrow 0$
the QCS (5.2) converts into the classical associativity condition for the
structure constants (5.1) [12].

We note that for the first time the system (5.2) has been derived in [37]
within a different context as the component-wise \ version of the central
system for the currents associated with the KP hierarchy. It was shown in
[37] that it encodes \ a complete algebraic information about the KP
hierarchy. We will demonstrate this in a little bit different manner.

Similar to the coisotropic case [12] there are, at least, two ways to decode
information contained in the QCS (5.2). First approach \ is to choose first
an appropriate parametrization of $H_{k}^{j}$ . As in the classical case
[12] we introduce the functions u,v and w \ defined by the formulae

\begin{equation}
H_{1}^{1}=-\frac{1}{2}u,H_{2}^{1}=-\frac{1}{3}v,H_{3}^{1}=-\frac{1}{4}w+%
\frac{1}{8}u^{2}.
\end{equation}

From the QCS (5.2) one gets

\begin{eqnarray*}
H_{1}^{2} &=&2H_{2}^{1}+\hbar \frac{\partial H_{1}^{1}}{\partial x^{1}}, \\
H_{1}^{3} &=&3H_{3}^{1}+\hbar \frac{\partial \left(
H_{2}^{1}+H_{1}^{2}\right) }{\partial x^{1}}, \\
H_{2}^{2} &=&-\hbar \frac{\partial H_{1}^{1}}{\partial x^{2}}%
-H_{3}^{1}+H_{1}^{3}+\left( H_{1}^{1}\right) ^{2}.
\end{eqnarray*}
Hence
\begin{eqnarray}
H_{1}^{2} &=&-\frac{2}{3}v-\frac{\hbar }{2}u_{x_{1}}, \nonumber \\
H_{1}^{3} &=&-\frac{3}{4}w+\frac{3}{8}u^{2}-\hbar v_{x_{1}}-\frac{\hbar ^{2}%
}{2}u_{x_{1}x_{1},} \nonumber \\
H_{2}^{2} &=&\frac{1}{2}u^{2}-\frac{1}{2}w+\frac{\hbar }{2}u_{x_{2}}-\hbar
v_{x_{1}}-\frac{\hbar ^{2}}{2}u_{x_{1}x_{1}}
\end{eqnarray}

Substituting these expressions into the first exactness conditions (5.4),i.e.
\begin{equation}
\frac{\partial H_{1}^{1}}{\partial x^{2}}-\frac{\partial H_{1}^{2}}{\partial
x^{1}}=0,\frac{\partial H_{2}^{1}}{\partial x^{2}}-\frac{\partial H_{2}^{2}}{%
\partial x^{1}}=0,\frac{\partial H_{1}^{1}}{\partial x^{3}}-\frac{\partial
H_{1}^{3}}{\partial x^{1}}=0
\end{equation}
and eliminating w, one gets the equations
\begin{eqnarray}
u_{x_{3}}-\frac{\hbar ^{2}}{4}u_{x_{1}x_{1}x_{1}}-\frac{3}{4}\left(
u^{2}\right) _{x_{1}}-\varphi _{x_{2}} &=&0, \nonumber \\
u_{x_{2}}-\frac{4}{3}\varphi _{x_{1}} &=&0
\end{eqnarray}
where $\varphi =v+\frac{3}{4}\hbar u_{x_{1}}.$ It is the famous
Kadomtsev-Petviashvili equation (see e.g. [33-35]). Using higher equations
(5.2) and (5.4), one in a similar manner obtains the higher KP equations and
the whole KP hierarchy.

In the limit $\hbar \rightarrow 0$ equation (5.9) is reduced to the
dispersionless KP equation while at the stationaly case $u_{x_{3}}=0$ one
gets the Boussinesq equation (4.9) at $\epsilon =0$ with $A=-u,\varphi =-D-%
\frac{\hbar }{4}A_{x_{1}}.$

Another way to deal with the QCS (5.2) is to solve first all exactness
conditions. \ One of them is given by (5.4). It implies the existence of the
functions $F_{k}$ such that

\begin{equation}
H_{k}^{j}=\frac{\partial F_{k}}{\partial x^{j}}, \qquad
j,k=1,2,3...
\end{equation}
The system (5.2) , in addition, contains another exactness type condition.
Indeed, as it was shown in [37], equations (5.2) lead to the following
constraint system
\begin{eqnarray}
\hbar \frac{\partial }{\partial x^{j}}\left(
\sum_{k=1}^{n-1}H_{n-k}^{k}\right)
+nH_{n}^{j}=H_{n}^{j}+\sum_{k=1}^{n-1}\left(
H_{k}^{j+n-k}-H_{j+n-k}^{k}\right)
+ \nonumber \\ +
\sum_{l=1}^{j-1}\sum_{k=1}^{n-1}H_{l}^{n-k}H_{k}^{j-l}.
\end{eqnarray}
The r.h.s. of (5.11) is symmetric  in the indices j and n. Hence

\begin{equation}
\hbar \frac{\partial }{\partial x^{j}}\left(
\sum_{k=1}^{n-1}H_{n-k}^{k}\right) +nH_{n}^{j}=\hbar \frac{\partial }{%
\partial x^{n}}\left( \sum_{k=1}^{j-1}H_{j-k}^{k}\right) +jH_{j}^{n}.
\end{equation}

Substitution of (5.10) into equations (5.12) gives the exactness conditions
\begin{equation}
\frac{\partial }{\partial x^{j}}\left( nF_{n}+\hbar \sum_{k=1}^{n-1}\frac{%
\partial F_{n-k}}{\partial x^{k}}\right) =\frac{\partial }{\partial x^{n}}%
\left( jF_{j}+\hbar \sum_{k=1}^{j-1}\frac{\partial F_{j-k}}{\partial x^{k}}%
\right) , \qquad j,n=1,2,3...
\end{equation}

\textbf{Proposition 5.2 }

\begin{equation}
H_{k}^{j}=\frac{1}{\hbar }\text{P}_{k}\left( -\hbar \widetilde{\partial }%
\right) F_{x_{j}}, \qquad  j,k=1,2,3,...
\end{equation}
where $P_{k}\left( -\hbar \widetilde{\partial }\right) \doteqdot P_{k}\left(
-\hbar \frac{\partial }{\partial x^{1}},-\frac{1}{2}\hbar \frac{\partial }{%
\partial x^{2}},-\frac{1}{3}\hbar \frac{\partial }{\partial x^{3}}%
,...\right) $ and $P_{k}\left( t_{1},t_{2},t_{3},...\right) $ are Schur
polynomials.

Proof. Equations (5.13) imply the existence of a function F such that

\begin{equation}
jF_{j}+\hbar \sum_{k=1}^{j-1}\frac{\partial F_{j-k}}{\partial x^{k}}%
=-F_{x_{j}}.
\end{equation}
Resolving (5.15) recurrently, one gets
\begin{eqnarray*}
F_{1} &=&-F_{x_{1}}, \\
2F_{2} &=&-F_{x_{2}}+\hbar F_{x_{1}x_{1},} \\
3F_{3} &=&-F_{x_{3}}+\frac{3}{2}\hbar F_{x_{1}x_{2}}-\frac{1}{2}\hbar
^{2}F_{x_{1}x_{1}x_{1}}
\end{eqnarray*}
and so on. The compact form of these relations is $F_{k}=\frac{1}{\hbar }%
P_{k}\left( -\hbar \widetilde{\partial }\right) F$ where Schur
polynomials are defined , as usual, by the generating formula
$\exp \left( \sum_{k=1}^{\infty }\lambda ^{k}t_{k}\right)
=\sum_{k=0}^{\infty }\lambda ^{k}P_{k}\left( \mathbf{t}\right) .$
Then, in virtue of (5.10) , one has (5.14)$\square $ .

Substitution of the expressions (5.14) for $H_{k}^{j}$ into the QCS (5.2)
gives the infinite system of differential equations,bilinear in F. The
simplest of them is

\begin{equation}
\frac{4}{3}F_{x_{1}x_{3}}-\frac{\hbar ^{2}}{3}F_{x_{1}x_{1}x_{1}x_{1}}-2%
\left( F_{x_{1}x_{1}}\right) ^{2}-F_{x_{2}x_{2}}=0.
\end{equation}

In terms of the function $\tau =\exp F$ the above equation and equations
(5.2) with

\begin{equation}
H_{k}^{j}=\frac{1}{\hbar }\text{P}_{k}\left( -\hbar \widetilde{\partial }%
\right) \frac{\tau _{x_{j}}}{\tau }, \qquad  j,k=1,2,3,...
\end{equation}
are nothing but the famous bilinear Hirota equations for the KP hierarchy (
see e.g. [34-36]). Hence, the function $\tau $ is the celebrated KP $\tau $%
-function.

Thus, any KP $\tau $-function defines weakly (non)associative quantum
deformations of the structure constants (5.1) for the infinite-dimensional
algebra by the formula (5.17) and Hirota bilinear equations. Quantum anomaly
for these deformations is given by (5.5) with

\begin{equation}
\left( A\right) _{k}^{n}=\frac{1}{\hbar }\sum_{l=n+1}^{\infty }P_{l-n}\left(
-\hbar \widetilde{\partial }\right) \frac{\tau _{x_{k}}}{\tau }dx^{l}.
\end{equation}

At last, in the limit $\hbar \rightarrow 0$ all the above formulae are
reduced to those for coisotropic deformations [12]. We empnasize that
quantum and isotropic deformations represent different deformations of the
same structure constants (5.1).

For concrete solutions of the KP hierarchy certain structure constants may
remain undeformed and components of quantum anomaly may vanish. For example,
the function F given by (4.14) with $\epsilon =0$ is the solution of
equation (5.16) too. For this solution all $H_{k}^{j}$ with j$\geq 3$ vanish
as well as $A_{k}^{n}=0$ for n,k$\geq 3$ . \ One soliton solution of the KP
equation corresponds to $\tau =1+\exp \left[ k\left(
x^{1}+px^{2}+qx^{3}\right) \right] $ where $q=\frac{\hbar ^{2}}{4}k^{2}+%
\frac{3}{4}p^{2}$ and k,p are arbitrary constants ( see e.g. [33-36]). For
this soliton deformation $H_{k}^{j}=0$ at j$\geq 4$ and $A_{k}^{n}=0$ for k,n%
$\geq 4.$

A quasi-triangular structure of the constants (5.1) allows us to rewrite
equations (2.5) in the equivalent form

\begin{equation}
\left( p_{n}-p_{1}^{n}-\sum_{m=1}^{n-2}u_{nm}\left( x\right)
p_{1}^{m}-u_{n0}p_{0}\right) \left| \Psi \right\rangle =0, \qquad
  n=1,2,3,...
\end{equation}
where the coefficients $u_{nm}$ are the certain functions of $H_{k}^{j}$. \
For example, $%
u_{20}=-2H_{1}^{1},u_{31}=-3H_{1}^{1},u_{30}=H_{2}^{1}+H_{1}^{2}+2\frac{%
\partial H_{1}^{1}}{\partial x^{1}}.$

In the coordinate representation \ equations (2.24) due to (2.25) take the
form

\begin{eqnarray}
-\hbar ^{2}\frac{\partial ^{2}\widetilde{\Psi }}{\partial x^{j}\partial x^{k}%
}+\hbar \frac{\partial \widetilde{\Psi }}{\partial x^{j+k}}+\hbar
\sum_{l=1}^{j-1}H_{j-l}^{k}\frac{\partial \widetilde{\Psi }}{\partial x^{l}}%
+\hbar \sum_{l=1}^{k-1}H_{k-l}^{j}\frac{\partial \widetilde{\Psi }}{\partial
x^{l}}+\left( H_{k}^{j}+H_{j}^{k}\right) \widetilde{\Psi }%
=0, \nonumber \\
 \qquad \qquad \qquad j,k=1,2,3,... \qquad \qquad
\end{eqnarray}

The system of linear equations (5.20) is equivalent to the standard set of
auxiliary linear problems for the KP hierarchy

\begin{equation}
\hbar \frac{\partial \widetilde{\Psi }}{\partial x^{n}}=\hbar ^{n}\frac{%
\partial ^{n}\widetilde{\Psi }}{\left( \partial x^{1}\right) ^{n}}%
+\sum_{m=0}^{n-2}\hbar ^{m}u_{nm}\left( x\right) \frac{\partial ^{m}%
\widetilde{\Psi }}{\left( \partial x^{1}\right) ^{m}}
\end{equation}
that is the coordinate representation of equations (5.19).

To get a standard form of the above formulae with a spectral parameter z one
considers a formal Laurent series $H^{\left( j\right) }\left( x,z\right)
\doteqdot \sum_{1}^{\infty }z^{-k}H_{k}^{j}.$ In virtue of (5.15) one has

\begin{equation}
H^{\left( j\right) }=\frac{1}{\hbar }\frac{\partial }{\partial x^{j}}\left\{
\left( \exp \left( -\hbar \sum_{n=1}^{\infty }\frac{z^{-n}}{n}\frac{\partial
}{\partial x^{n}}\right) -1\right) F\right\} =\frac{1}{\hbar }\frac{\partial
}{\partial x^{j}}\log \left( \frac{\tau \left( x-\left[ z^{-1}\right]
\right) }{\tau \left( x\right) }\right)
\end{equation}
where $x-\left[ z^{-1}\right] \doteqdot \left( x^{1}-\frac{1}{z},x^{2}-\frac{%
1}{2z^{2}},x^{3}-\frac{1}{3z^{3}},...\right) $ is the Miwa shift [36].
Introducing the wave function $\chi $ by $H^{\left( j\right) }\doteqdot
\frac{1}{\hbar }\frac{\partial \log \chi }{\partial x^{j}}$ ,one obtains

\begin{equation}
\chi \left( x,z\right) =\frac{\tau \left( x-\left[ z^{-1}\right] \right) }{%
\tau \left( x\right) }
\end{equation}
that reproduces the standard form of the dressed KP wave-function

\begin{equation}
\widetilde{\Psi }\left( x,z\right) =\exp \left( \sum_{n=1}^{\infty
}z^{n}x^{n}\right) \chi \left( x,z\right) =\exp \left( \sum_{n=1}^{\infty
}z^{n}x^{n}\right) \frac{\tau \left( x-\left[ z^{-1}\right] \right) }{\tau
\left( x\right) }
\end{equation}
in terms of the $\tau -$ function [36]. For more details see [37].

\ It is well-known that the stationary reductions of the KP hierarchy \ \
give rise to the Gelfand-Dickey \ hierarchies ( see e.g. [33-35]). \ At the
same time one can show that the stationarity constraint $\frac{\partial
H_{k}^{j}}{\partial x^{N}}=0$ converts the infinite-dimensional polynomial
algebra into the finite-dimensional one. \ So, stationary solutions of the
KP hierarchy provide us with the weakly (non)associative quantum
deformations of the finite-dimensional algebras. The Boussinesq deformation
(4.8)-(4.12) is the simplest example. For the general Gelfand-Dickey case
see also [18].

Finally, we would like to note that the quantum deformations of algebras
obtained by the process \ of gluening [12] of \ N algebras of the type (5.1)
are described by the N-component KP hierarchy.

At last, in the limit $\hbar \rightarrow 0$ all the above formulae
are reduced to those for coisotropic deformations [12]. We
empnasize that quantum and isotropic deformations represent
different \ deformations of the same structure constants (5.1).

\bigskip

\textbf{6. Conclusion}

The approach presented in the paper can be extended in different directions.
For instance, \ the basic idea of identification of the elements \textbf{P}$%
_{j}$ of the basis and deformations parameters x$^{j}$ with the elements of
the Heisenberg algebra can be applied to other type of algebras.

We note also that the formula (2.10) gives a simple realization for
previously discussed idea \ of geometrical interpretation of the associator
for a algebra as a curvature tensor (see e.g. [38] ). It suggests \ a
natural generalization of quantum deformations to nonassociative algebras.

\textbf{Acknowledgement.} The author is very grateful to Franco Magri for
numerous fruitful discussions.

\bigskip

\textbf{References}

\bigskip

1. Gerstenhaber M., On the deformation of rings and algebras, Ann.
Math., \textbf{79}, 59-103 (1964).

2. Gerstenhaber M., On the deformation of rings and algebras. II,
Ann. Math., \textbf{84}, 1-19 (1966).

3. Witten E., On the structure of topological phase of two-dimensional
gravity, Nucl. Phys., \textbf{B 340}, 281-332 (1990).

4. Dijkgraaf R., Verlinde H. and Verlinde E., Topological strings in d%
\TEXTsymbol{<}1, Nucl. Phys., \textbf{\ B 352,} 59-86 (1991).

5. Dubrovin B., Integrable systems in topological field theory,
Nucl. Phys., \textbf{B 379}, 627-689 (1992).

6. Dubrovin B., \ Geometry of 2D topological field theories, Lecture Notes
in Math., \textbf{1620}, 120-348 (1996), Springer, Berlin.

7. Hertling C. and Manin Y.I., Weak Frobenius manifolds, Int. Math. Res.
Notices, \textbf{6}, 277-286 (1999).

8. Manin Y.I., F-manifolds with flat structure and Dubrovin's duality, Adv.
Math., \textbf{198}, 5-26 (2005).

9. Manin Y.I., \textit{Frobenius manifolds, quantum cohomology and moduli
spaces}, AMS, Providence, 1999.

10. Hertling C.,\textit{\ Frobenius manifolds and moduli spaces for
singularities}, Cambridge Univ. Press, 2002.

11. Hertling C. and Marcoli M. (Eds.), \ \textit{Frobenius manifolds,
quantum cohomology and singularities}, \ Aspects of Math., \textbf{E36},
2004, \ \ \ \ Friedr. Vieweg \& Sohn, Wiesbaden.

12. Konopelchenko B.G. and \ Magri F., \ Coisotropic deformations of
associative algebras and dispersionless integrable hierarchies,
Commun. Math. Phys., \textbf{274}, 627-658 (2007).

13. Konopelchenko B.G. and Magri F., Dispersionless integrable equations as
coisotropic deformations: extensions and reductions, Theor. Math. Phys.,
\textbf{151}, 803-819 (2007).

14. Konopelchenko B.G. and Magri F., Coisotropic deformations of
associative algebras and integrable systems, to appear.

15. Dirac P.A.M., \textit{Lectures on quantum mechanics,} \ Yeshiva Univ.,
New York, 1964.

16. Eisenhart L.P.,\textit{\ Riemann geometry}, Princeton Univ. , 1926.

17. Willmore T.J., \textit{Riemann geometry}, \ Oxford Science Publ.,
Claderon Press, New York, 1993.

18. Givental A., $\ A_{n-1}$ singularities and nKdV hierarchies, Mosc. Math.
J., \textbf{3}, 475-505 (2003).

19. Coates T. and Givental A., Quantum Riemann-Roch, Lefschetz and Serre,
Ann. Math., \textbf{165}, 15-53 (2007).

20. Givental A., Gromov-Witten invariants and quantization of quadratic
Hamiltonians, Mosc. Math. J., \textbf{1}, 551-568 (2001).

21. Eliashberg Y., Symplectic field theory and its applications, Plenary
Lectures and Ceremonies, ICM 2006, European Math. Soc., 2007, pp. 217-246.

22. Cattaneo A.S. and Felder G., Relative formality theorem and quantization
of coisotropic manifolds, Adv. Math., \textbf{208}, 521-548 (2007).

23. Losev A. and Manin Y.I., Extended modular operads, in : \textit{%
Frobenius manifolds, quantum cohomology and singulatities} (C. Hertling and
M. Marcoli M. , Eds.), Aspects of Math., \textbf{E36}, 181-211 (2004).

24. Dubrovin B., Geometry and integrability of topological-antitopological
fusion, Commun. Math. Phys., \textbf{152}, 539-564 (1993).

25. Givental A. and Kim B., Quantum cohomology of flag manifolds and Toda
lattices, Commun. Math. Phys., \textbf{168}, 609-642 (1995).

26. Kito H., On Hessian structures of the Euclidean space and hyperbolic
space, Osaka J. Math., \textbf{36}, 51-62 (1999).

27. Dubrovin B., \ On almost duality for Frobenius manifolds, in:\textit{\
Geometry, topology and mathematical physisc, }\ Amer. Math. Soc. Trans.,
\textbf{212}, 75-132 (2004), AMS, Providence.

28. Nomizu K. and Sasaki T., \textit{Affine differential geometry}, \
Cambridge Univ.Press, 1994.

29. Ferapontov E.V., Hypersurfaces with flat centroaffine metric and
equations of associativity, Geom. Dedicata, \textbf{103}, 33-49 (2004).

30. Mokhov O.T., Theory of submanifolds, associativity equations in 2D
topological quantum field theory and Frobenius manifolds, Theor. Math.
Phys., \textbf{152}, 1183-1190 (2007).

31. Whitham G.B.,\textit{\ Linear and nonlinear waves}, Wiley/ Interscience,
New York, 1974.

32. Zakharov V.E., On stochastization of one-dimensional chain of nonlinear
oscillators, Sov. Phys. JETP, \textbf{35}, 908-914 (1974).

33. Novikov S.P., Manakov S.V., Pitaevski L.P. and Zakharov V.E.,\textit{\
Theory of solitons.The inverse problem method, }Plenum, New\textit{\ }York%
\textit{,1984.}

34. Ablowitz M.J. and Clarkson P.A., \textit{Solitons, nonlinear evolution
equations and inverse scattering,} Cambridge Univ.Press, 1991.

35. Konopelchenko B.G.,\textit{\ Introduction to multidimensional integrable
equations}, Plenum Press, New York and London, 1992.

36. Jimbo M. and Miwa T., Solitons and infinite-dimensional Lie algebras,
Publ. RIMS, Kyoto Univ., \textbf{19}, 943-1001 (1983).

37. Falgui G., Magri F. and Pedroni M., Bihamiltonian geometry, Darboux
covering and linearization of the KP hierarchy, Commun. Math. Phys., \textbf{%
197}, 303-324 (1998); Casati P., Falqui G., Magri F. and Pedroni M., The KP
theory revisited. IV. KP equations, dual KP equations, Baker-Akhiezer and $%
\tau $ functions, Preprint SISSA/5/96/FM, Trieste, Italy, 1996.

38. Ionescu L.M., Nonassociative \ algebras: a framework for differential
geometry, Intern. J. Math. and Math. Sciences, \textbf{60}, 3777-3795 (2003).

\bigskip

\end{document}